\documentclass[aps,prb,twocolumn,notitlepage,showpacs,superscriptaddress,am]{revtex4-2}%
\usepackage{graphicx}
\usepackage{amsmath}
\usepackage{amssymb}
\usepackage{mhchem}
\usepackage{color}
\usepackage{multirow}
\usepackage{amsfonts}%
\usepackage[breaklinks=true,colorlinks=true,linkcolor=blue,urlcolor=blue,citecolor=blue]{hyperref}

\providecommand{\U}[1]{\protect\rule{.1in}{.1in}}

\begin{document}
\title{High-quality single crystals of the kagome metals \ce{Ni3In} and \ce{Ni3Sn} grown from Pb flux}

\author{F. Garmroudi}
\email{fgarmroudi@lanl.gov}
\affiliation{Materials Physics Applications--Quantum, Los Alamos National Laboratory, Los Alamos, New Mexico 87545, USA}
\author{J. Coulter}
\email{jcoulter@flatironinstitute.org}
\affiliation{Center for Computational Quantum Physics, Flatiron Institute, New York 10010, USA}
\author{C.\,S.\,T. Kengle}
\affiliation{Materials Physics Applications--Quantum, Los Alamos National Laboratory, Los Alamos, New Mexico 87545, USA}
\author{J.\,D. Thompson}
\affiliation{Materials Physics Applications--Quantum, Los Alamos National Laboratory, Los Alamos, New Mexico 87545, USA}
\author{E.\,D. Bauer}
\affiliation{Materials Physics Applications--Quantum, Los Alamos National Laboratory, Los Alamos, New Mexico 87545, USA}
\author{S.\,M. Thomas}
\affiliation{Materials Physics Applications--Quantum, Los Alamos National Laboratory, Los Alamos, New Mexico 87545, USA}
\author{P.\,F.\,S. Rosa}
\email{pfsrosa@lanl.gov}
\affiliation{Materials Physics Applications--Quantum, Los Alamos National Laboratory, Los Alamos, New Mexico 87545, USA}


\begin{abstract}
The bilayer kagome metal \ce{Ni3In} has recently attracted attention due to the presence of a flat band located near the Fermi level, which has been associated with unconventional thermodynamic and electronic transport properties [Ye et al., Nat. Phys. 20, 610--614 (2024)]. However, further investigation of the intrinsic properties of this system has been hindered by the lack of large, high-quality single crystals. Here, we report the successful growth of \ce{Ni3(In,Sn)} single crystals using a Pb-flux technique. By optimizing the growth conditions, competing binary phases can be effectively suppressed, enabling the synthesis of single crystals with dimensions reaching several millimeters. We compare the physical properties of our Pb-flux-grown crystals to previously reported samples prepared by iodine-assisted chemical vapor transport and molecular beam epitaxy as well as to first-principles resistivity calculations. We find a significantly lower electrical resistivity in our crystals, in excellent agreement with calculations of resistivity from electron-phonon scattering, a sizeable non-saturating magnetoresistance, and a reduced Sommerfeld coefficient and magnetic susceptibility compared to previous experimental findings, which are likely related to differences in the Fermi level position. Our results establish Pb-flux growth as a reliable route for obtaining large single crystals of the bilayer kagome metals \ce{Ni3(In,Sn)} that are suitable for further thermodynamic and spectroscopic investigations of their intrinsic electronic and magnetic properties.
\end{abstract}

\maketitle

\label{intro}
Kagome metals have emerged as a fertile platform for exploring how lattice geometry, band topology, and electronic correlations intertwine to give rise to new states of matter \cite{ye2018massive,ortiz2019new,thomas2025unusual,riedel2025magnetic,cheng2026interwoven,di2026kagome}. The term “kagome” refers to a two-dimensional network of corner-sharing triangles in the crystal structure of these materials. This structural motif naturally produces Dirac crossings and van Hove singularities at specific points in momentum space as well as dispersionless flat bands (FBs) arising from destructive interference of nearest-neighbor hopping paths on the geometrically frustrated lattice \cite{kang2020topological,liu2020orbital,meier2020flat,kang2020dirac}. These features make kagome metals especially susceptible to symmetry-breaking instabilities such as charge order, magnetism, and superconductivity, while also providing a natural setting for studying correlation-driven topological phases in bulk materials \cite{chen2014anomalous,mazin2014theoretical,xu2015intrinsic,ye2018massive,tan2021charge,liu2024superconductivity,
park2021electronic}.

A major challenge is to identify systems in which the flat bands reside at or very near the Fermi energy $E_\text{F}$, allowing them to directly influence the low-energy electronic properties. The bilayer kagome metal \ce{Ni3In} (Fig.\,1a) has recently emerged as a particularly promising candidate, as it hosts a flat band nearly pinned to $E_\text{F}$ (Fig.\,1b) together with a Dirac nodal ring \cite{ye2024hopping}. Initial investigations \cite{ye2024hopping} revealed several unconventional properties: (i) a large Sommerfeld coefficient $\gamma\approx 51.6\,$mJ\,mol$_\text{f.u.}^{-1}$\,K$^{-2}$, at least three times larger than the value expected from the density of states (DOS) calculated by density functional theory (DFT); (ii) an electrical resistivity that remains linear in temperature down to low temperatures $(T_\text{FL}\approx 1.5\,$K) reminiscent of the “strange-metal” behavior observed in correlated high-temperature superconductors and heavy-fermion compounds in the proximity of quantum critical points; and (iii) an unusually large Kadowaki-Woods ratio $A/\gamma^2\approx 10\,\mu \Omega$\,cm\,mol$^2$\,K$^2$\,J$^{-2}$, exceeding the values found in elemental Ni and conventional transition metals by more than two orders of magnitude. Here, $A$ denotes the coefficient of the quadratic low-temperature resistivity contribution, $\rho-\rho_0=AT^2$, below 1.5\,K. 

Motivated by these observations, a heavy-fermion-like ground state has been proposed for \ce{Ni3In}, in which the Ni $d_{ij}$-derived compact localized states mimic the role analogous to the localized $f$ orbitals in conventional heavy-fermion systems and hybridize with the dispersive conduction bands through a Kondo-like mechanism \cite{ye2024hopping,mahankali2025correlated}. Temperature-dependent polarized Raman scattering showed frequency and linewidth renormalizations below $T^*\approx 50\,$K, which has been interpreted as a possible signature of incoherent--coherent Kondo crossover \cite{gim2023fingerprints}. Lastly, recent scanning tunneling microscopy measurements \cite{souza2026origin} further revealed a temperature-dependent Fano-like zero-bias peak--dip structure sensitive to temperature and magnetic field, underscoring the importance of electronic correlations and suggesting an emergent many-body character in the low-temperature state of \ce{Ni3In}.

\begin{figure*}[t!]
\centering
 \includegraphics[width=0.95\textwidth]{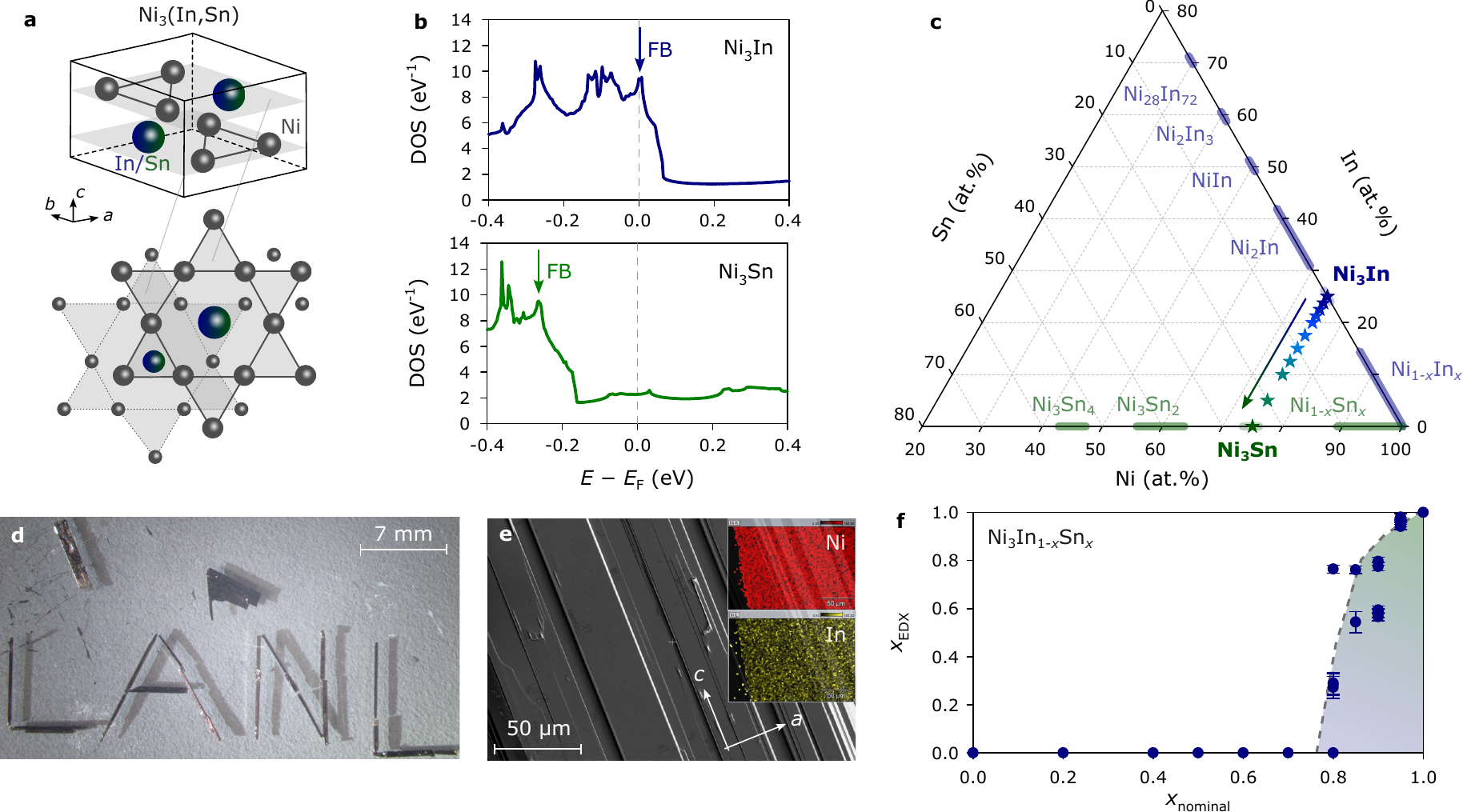}
\caption{\textbf{a} Hexagonal (P6$_3$/mmc) crystal structure of \ce{Ni3In} and \ce{Ni3Sn}, where Ni atoms form $AB$ stacked kagome bilayers and In/Sn atoms occupy the center of the hexagons.  \textbf{b} Densities of states of \ce{Ni3In} and \ce{Ni3Sn} from density functional theory. While the former has a flat band (vertical arrow) exactly at $E_\text{F}$, the Fermi level is above the FB for the latter, allowing continous tuning of the FB edge with respect to $E_\text{F}$ in \ce{Ni3In_{1-x}Sn_x}. \textbf{c} Known phases in the ternary Ni--In--Sn system. \ce{Ni3In} and \ce{Ni3Sn} form a full solid solution \ce{Ni3In_{1-x}Sn_x}, various compositions of which were investigated in this work. \textbf{d} Large \ce{Ni3In} single crystals (up to 7--8\,mm long and up to 1--2\,mm wide) obtained from a starting composition of 1\,mol \ce{Ni3In_{0.5}Sn_{0.5}} in 25\,mol Pb (see Experimental Methods for details). \textbf{e} Scanning electron microscopy image of an unpolished \ce{Ni3In} single-crystal surface used for transport experiments. Insets display EDX elemental maps of Ni and In, revealing a homogeneous distribution and no detectable Pb inclusions. \textbf{f} Sn concentration in \ce{Ni3In_{1-x}Sn_x} single crystals determined by energy-dispersive X-ray spectroscopy (EDX) as a function of nominal Sn content.
}
\label{Fig1}
\end{figure*} 

Beyond their fundamental interest, Sn-substituted \ce{Ni3In_{1-x}Sn_x} kagome metals may also provide a promising route toward ultrahigh thermoelectric performance, potentially rivaling or even surpassing commercial \ce{Bi2Te3}-based semiconductors \cite{garmroudi2025topological} as well as other state-of-the-art room-temperature thermoelectrics \cite{mao2019high,zhao2024plasticity,garmroudi2022anderson,garmroudi2025decoupled,garmroudi2025recent}. Tuning the Fermi level through Sn substitution such that the flat-band states lie slightly below $E_\text{F}$
can produce a pronounced electron--hole asymmetry in carrier transport: hole-like excitations are strongly scattered and effectively filtered out, while electron-like carriers remain comparatively mobile \cite{garmroudi2024high,garmroudi2025topological}. This intrinsic asymmetric energy-filtering mechanism has been shown to generate unusually large Seebeck coefficients in otherwise metallic systems \cite{garmroudi2023high,garmroudi2025energy,garmroudi2025topological}.

A bottleneck for the further investigation and understanding of \ce{Ni3In}-based kagome metals has been the realization of large, high-quality single crystals. Previous studies relied on small crystals (typically only a few hundred microns in size) obtained by iodine-assisted recrystallization \cite{ye2024hopping}, molecular-beam-epitaxy-grown thin films \cite{han2024molecular} with relatively low residual resistivity ratios (RRR), or polycrystalline samples \cite{garmroudi2024high,garmroudi2025topological} that obscure the intrinsic anisotropy of the electronic structure. In this work, we report the growth and physical properties of large \ce{Ni3In_{1-x}Sn_x} single crystals with improved sample quality, which were synthesized using a Pb-flux method.

The remainder of this paper is organized as follows. In \ref{growth}, we discuss the growth conditions and key considerations required to obtain large \ce{Ni3In_{1-x}Sn_x} single crystals. In \autoref{transport}, we present and analyze the electrical resistivity of \ce{Ni3In} and \ce{Ni3Sn} over a broad temperature range and compare the results with DFT-based transport calculations and previous experiments. Finally, in \autoref{magnetism}, we present measurements of the magnetic susceptibility and specific heat and compare our results with those reported previously for crystals grown by iodine-assisted recrystallization.

\section{Results and Discussion}
\subsection{Pb flux growth of \ce{Ni3In_{1-x}Sn_x}}
\label{growth}
Owing to the complexity of the binary Ni--In phase diagram, which contains multiple intermetallic phases with high melting temperatures, the growth of \ce{Ni3In} single crystals from self flux is not feasible as this phase does not coexist with the liquid phase over any composition range \cite{durussel1997binary}. The same is true for \ce{Ni3Sn} \cite{schmetterer2007new} and most likely for the continuous solid solution \ce{Ni3In_{1-x}Sn_x} (Fig.\,1c). It is therefore necessary to identify an alternative flux material. We find that Pb is particularly suitable, as it forms no competing binary phases with Ni, In, or Sn and exhibits good solubility for all constituent elements at moderately elevated temperatures \cite{pomianek1986thermodynamische,nash1987ni,karakaya1988pb,nabot1987pb}. For example, the liquidus temperature in the binary Ni--Pb system at 90 at.\% Pb is only about 1030\,$^\circ\mathrm{C}$, and one may expect it to be even lower in the pseudobinary \ce{Ni3In}--Pb system since \ce{Ni3In} melts/decomposes at a lower temperature than elemental Ni \cite{durussel1997binary}.

\begin{table}[b!]
\caption{Flux growth of \ce{Ni3In_{1-x}Sn_x} kagome metals -- nominal compositions of samples investigated in this work, maximum temperature ($T_\text{soak}$) during growth, observed phases, and composition from energy-dispersive x-ray spectroscopy.}
\label{tab:3x5}
\begin{tabular}{cccc}
\hline
composition \,\,& $T_\text{soak}$ ($^\circ$C)  \,\,& \,observed phases \,& $x_\text{EDX}$  \\ \hline

\ce{Ni3In} & 1000 & \ce{Ni2In}, \ce{Ni3In} & --  \\

\ce{Ni3In} & 800 & \ce{Ni3In}  & --  \\

\ce{Ni3In_{0.8}Sn_{0.2}} & 800 & \ce{Ni3In} & --  \\

\ce{Ni3In_{0.6}Sn_{0.4}} & 800 & \ce{Ni3In} & --  \\

\ce{Ni3In_{0.5}Sn_{0.5}} & 800 & \ce{Ni3In} & --  \\

\ce{Ni3In_{0.2}Sn_{0.8}} & 800 & \ce{Ni3In_{1-x}Sn_x} & 0.28  \\

\ce{Ni3In_{0.5}Sn_{1.33}} & 800 & \ce{Ni3Sn_{2}}, \ce{Ni3In} & -- \\

\ce{Ni3In_{0.5}Sn_{0.5}} & 1000 & \ce{Ni3In}, \ce{Ni2In} & --  \\

\ce{Ni3In_{0.3}Sn_{0.7}} & 1000 & \ce{Ni3In} & --  \\

\ce{Ni3In_{0.2}Sn_{0.8}} & 1000 & \ce{Ni3In_{1-x}Sn_x} & 0 \& 0.76  \\

\ce{Ni3In_{0.15}Sn_{0.85}} & 1000 & \ce{Ni3In_{1-x}Sn_x} & 0.54 \& 0.76  \\

\ce{Ni3In_{0.1}Sn_{0.9}} & 1000 & \ce{Ni3In_{1-x}Sn_x} & 0.58 \& 0.79  \\

\ce{Ni3In_{0.05}Sn_{0.95}} & 1000 & \ce{Ni3In_{1-x}Sn_x} & 0.96  \\

\ce{Ni3Sn} & 800 & \ce{Ni3Sn} & --  \\
\hline
\end{tabular}
\end{table}

\begin{figure*}[t!]
\centering
 \includegraphics[width=1\textwidth]{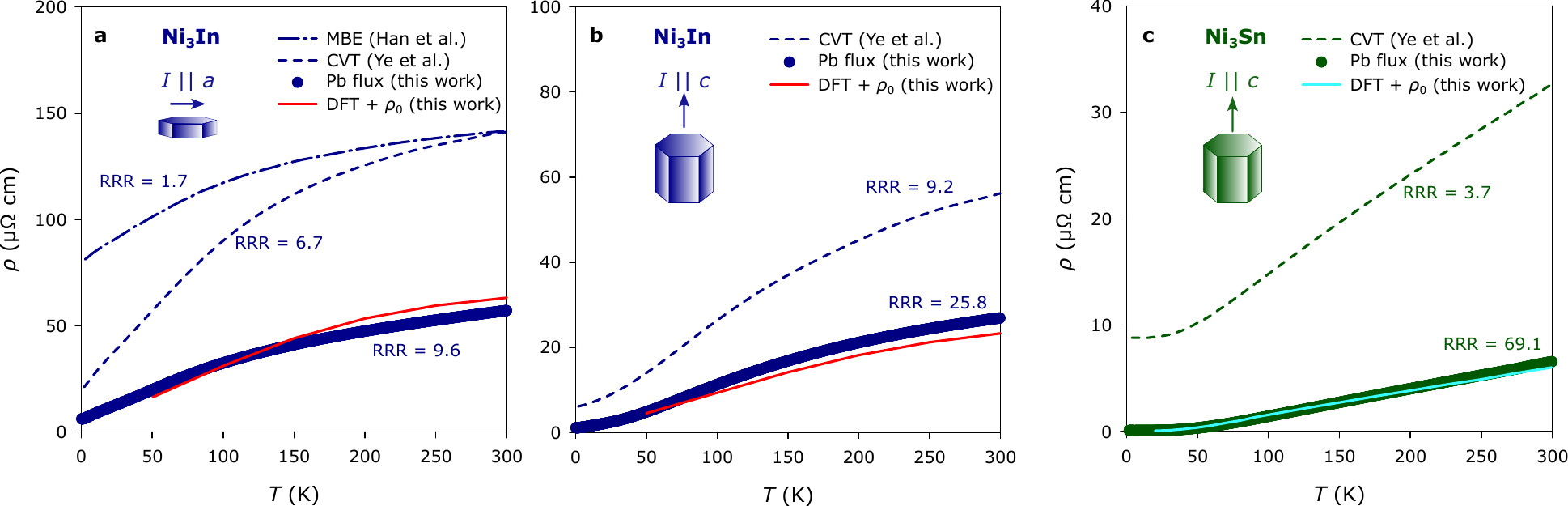}
\caption{\textbf{a} Temperature-dependent electrical resistivity of \ce{Ni3In} with current applied along the crystallographic $a$-axis (within the kagome planes) and \textbf{b} along the $c$-axis (perpendicular to the kagome planes). \textbf{c} Same as in \textbf{b}, but for \ce{Ni3Sn}. Single crystals grown from Pb flux (this work) exhibit significantly improved sample quality, as evidenced by much lower residual resistivities $\rho_0$ and higher residual resistivity ratios (RRR), compared to crystals grown by iodine-assisted chemical vapor transport (CVT) \cite{ye2024hopping} or molecular beam epitaxy (MBE) \cite{han2024molecular}. Solid lines represent first-principles DFT-based electron--phonon transport calculations performed within the Wigner transport formalism, including the experimentally determined residual resistivity.
}
\label{Fig2}
\end{figure*} 

We investigated a range of nominal \ce{Ni3In_{1-x}Sn_x} compositions (see Table\,1), all dissolved in 25\,mol of Pb flux. Starting with \ce{Ni3In}, we find that growth from 1000\,$^\circ\mathrm{C}$ leads to the simultaneous precipitation of a competing hexagonal \ce{Ni2In} phase, resulting in only few relatively small \ce{Ni3In} crystals. Lowering the growth temperature to 800\,$^\circ\mathrm{C}$ mitigates this problem and suppresses the formation of \ce{Ni2In}, which forms at higher temperatures than \ce{Ni3In}, yielding moderately sized single crystals reaching up to $\approx 2$\,mm along the $c$ axis and up to 200--300\,$\mu$m in the $ab$ plane. Interestingly, for nominal \ce{Ni3In_{1-x}Sn_x} compositions up to $x\approx 0.7-0.8$, no Sn is incorporated into the crystals, and stoichiometric \ce{Ni3In} crystals are obtained instead (Fig.\,1f). Adding extra Sn and changing the ratio 3\,Ni\,:\,1\,(In,Sn) further proved unsuccessful, as the formation of \ce{Ni3Sn2} is promoted.

Because \ce{Ni3Sn} forms at higher temperatures than \ce{Ni3In} \cite{durussel1997binary,schmetterer2007new}, we increased the soak temperature $T_\text{soak}$ back to 1000\,$^\circ\mathrm{C}$ in an attempt to enhance Sn incorporation. However, this did not resolve the discrepancy between nominal and actual Sn concentrations in the resulting crystals. Instead, we unexpectedly found that starting compositions with nominal Sn concentrations around $x=0.5$ produce exceptionally large single crystals of stoichiometric \ce{Ni3In}, reaching sizes up to $\approx 8$\,mm along the $c$ axis and up to 1--2 mm in the $ab$ plane (Fig.\,1d). We attribute this effect to the modified Ni\,:\,In ratio in the melt. The relatively increased Ni content likely suppresses the formation of the competing high-temperature \ce{Ni2In}-type phase, thereby enabling the growth of significantly larger \ce{Ni3In} crystals from the solution. This indicates that optimal growth of \ce{Ni3In} in Pb flux can be achieved for 6\,Ni\,:\,1\,In\,:\,50\,Pb. The growth of \ce{Ni3In_{1-x}Sn_x} has proven to be unexpectedly challenging and requires careful empirical optimization of the starting composition and verification of the composition by energy-dispersive x-ray spectroscopy. Exploring alternative flux materials may provide a route toward improved Sn incorporation into the crystals. In the following, we focus on the properties of the stoichiometric end members \ce{Ni3Sn} and, in particular, \ce{Ni3In}. For \ce{Ni3In}, all measurements were performed on single crystals grown from the batch with a starting composition 3\,Ni\,:\,0.5\,In\,:\,0.5\,Sn\,:\,25\,Pb and $T_\text{soak}=1000\,^\circ$C. The absence of any point defects (antisites, Sn or Pb doping) and any sizeable flux inclusions for these crystals was confirmed by single-crystal X-ray diffraction and energy-dispersive X-ray spectroscopy. These measurements are also essential for distinguishing \ce{Ni3In} from \ce{Ni2In}, as both phases form crystals with very similar morphologies.

\begin{figure}[t!]
\centering
 \includegraphics[width=0.35\textwidth]{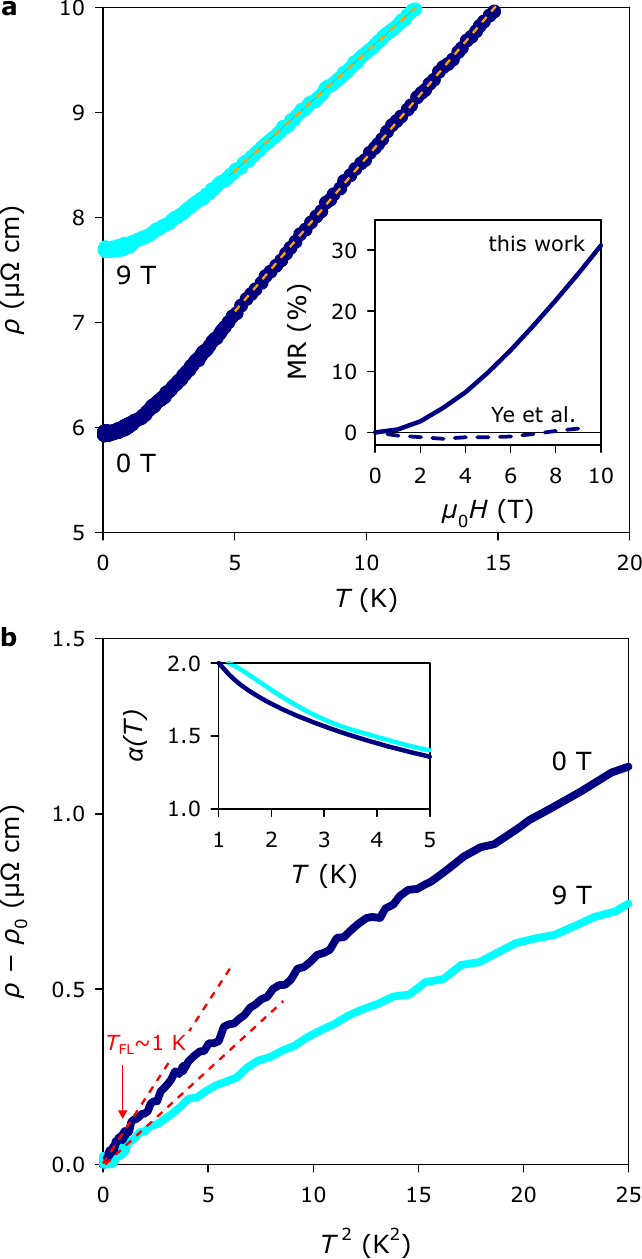}
\caption{\textbf{a} Low-temperature resistivity of \ce{Ni3In}, measured with current along the $a$-axis. A comparison of $\rho(T)$ at zero field and in an applied field $\mu_0H = 9\,$T along $b$ reveals a sizeable positive magnetoresistance and a non-Fermi-liquid temperature dependence down to very low temperatures in both cases (cf. panel \textbf{b}). Inset shows the field-dependent magnetoresistance $\text{MR} = [\rho(H)-\rho(0)]/\rho(0)$ at 2\,K. Crystals grown by iodine-assisted recrystallization \cite{ye2024hopping} display vanishingly small MR, whereas crystals grown from Pb flux (this work) display a sizeable nonsaturating MR reaching $\approx 30\,\%$ at 10\,T. \textbf{b} Both a plot of $\rho-\rho_0$ versus $T^2$ and the local resistivity exponent, $\alpha(T)=d\ln \left[\rho(T)-\rho_0\right]/d\ln T$ (inset), reveal a crossover to Fermi-liquid behavior, $\rho(T)-\rho_0=AT^2$, only below $T_\text{FL}\approx 1-1.2\,$K with $A\approx 0.08\,\mu\Omega$\,cm\,K$^{-2}$.
}
\label{Fig3}
\end{figure} 
\subsection{Electrical transport properties}
\label{transport}
Figure 2 compares the electrical resistivity of \ce{Ni3In} and \ce{Ni3Sn}, measured along the $a$ and $c$ directions for \ce{Ni3In} and, due to the smaller crystal dimensions, only along the $c$ direction for \ce{Ni3Sn}. These results are compared to previously reported data obtained from samples prepared using different synthesis techniques (chemical vapor transport \cite{ye2024hopping} and molecular beam epitaxy \cite{han2024molecular}), as well as to DFT-based \textit{ab initio} electron--phonon transport calculations performed in this work in the framework of the relaxation time approximation and Wigner transport formalism (RTA-WTE) \cite{cepellotti2021interband}, as implemented in the open-source package Phoebe \cite{cepellotti2022phoebe}. 

It is evident that the crystal quality of our samples is significantly improved compared to \cite{ye2024hopping} and \cite{han2024molecular}, particularly along the $c$-axis, as reflected by a substantially reduced residual resistivity $\rho_0$ and enhanced residual resistivity ratios $\text{RRR}=\rho_\text{300}/\rho_0$ (see also Fig.\,S2). $\mathrm{RRR}$ values reach up to 10, 26, and 69 for \ce{Ni3In}\,$\parallel\!a$, \ce{Ni3In}\,$\parallel\!c$, and \ce{Ni3Sn}\,$\parallel\!c$, respectively -- larger than those found previously. We attribute the difference in RRR between \ce{Ni3In}\,$\parallel\!a$ and $\parallel\!c$ to grain boundary scattering, as the crystals grow in columnar grains along $c$ (see Fig.\,1c and Fig.\,S3) and $a$-axis samples were cut perpendicular to the grain-growth direction. Both the absolute magnitude and the overall temperature dependence of the resistivity are in remarkable agreement with our DFT-based \textit{ab initio} electron--phonon transport calculations, which contain no free parameters ($\delta \rho_\text{calc-exp}<15\,$\%). This agreement suggests that electron--electron correlations are not dominant at high temperatures. We note that the temperature-dependent Raman scattering performed in Ref.\,\cite{gim2023fingerprints} showed frequency and linewidth renormalizations only below $T^*\approx 50\,$K. It is also interesting that, while $\rho(T)$ of the MBE-grown samples could be explained by merely shifting up $\rho_\text{DFT}$ by a different $\rho_0$, the resistivity of the CVT-grown samples appears to be larger by a multiplicative factor ($\approx$ 2.5). We speculate that this could be explained by a different Fermi level position, deeper within the FB region in the CVT-grown crystals, where the group velocities are smaller. Irrespective of the scattering mechanism, the resistivity depends inversely on the squared group velocities $v^2$, and $v^2$ can be expected to be strongly energy-dependent in \ce{Ni3In}, since $E_\text{F}$ is located close to a FB.

Finally, we point out the low room-temperature resistivity of \ce{Ni3Sn} along the $c$ axis, $\rho_{c,300} \approx 6.5\,\mu\Omega\,\mathrm{cm}$. It has been shown that the Fermi-surface anisotropy and low $\rho_c$ found in other kagome metals such as CoSn can suppress carrier--surface scattering, resulting in even lower $\rho$ than Cu when decreasing the thickness down to the nanoscale \cite{kumar2022ultralow}, which may make \ce{Ni3Sn} of interest for applications in electrical interconnects. 

To further investigate the strange-metal state of \ce{Ni3In} and search for possible ordered phases, we performed measurements down to dilution-refrigerator temperatures with the electrical current applied parallel to the $a$ axis. No superconducting or magnetic phase transitions were observed down to 60\,mK. Figure\,3 shows the temperature-dependent electrical resistivity along the $a$ axis, $\rho_a(T)$, measured in zero field and in an applied magnetic field of 9\,T. Contrary to previous findings \cite{ye2024hopping}, the strange-metal state and the low-temperature resistivity exponent (inset Fig.\,3b) are remarkably robust against magnetic fields in our crystals. Moreover, despite the sizeable difference in the absolute values of $\rho_a(T)$ between our crystals and those studied previously (Fig.\,2a), the non-Fermi-liquid behavior persists over a similar temperature range down to $T \approx 1\,$K. At the lowest temperatures, $\rho_a(T)-\rho_{a,0}$ can be well described by an $AT^2$ dependence. These observations demonstrate that the strange-metal behavior is neither driven by disorder nor strongly affected by the growth method, but is instead intrinsically connected to the unusual electronic structure within the kagome planes of \ce{Ni3In}. Finally, we note that the magnetoresistance in our crystals is substantially larger than that reported in Ref.\,\cite{ye2024hopping}, likely reflecting enhanced carrier mobilities in our samples and proximity of the Fermi level to the Dirac-like band crossing. As discussed before, it is possible that the Fermi level in the CVT-grown samples is shifted slightly away from the Dirac nodal ring and deeper into the flat-band region. Such a shift could also possibly account for the differences observed in the Sommerfeld coefficient of the specific heat and in the magnetic susceptibility, as discussed in the following section.

\begin{figure*}[t!]
\centering
 \includegraphics[width=0.95\textwidth]{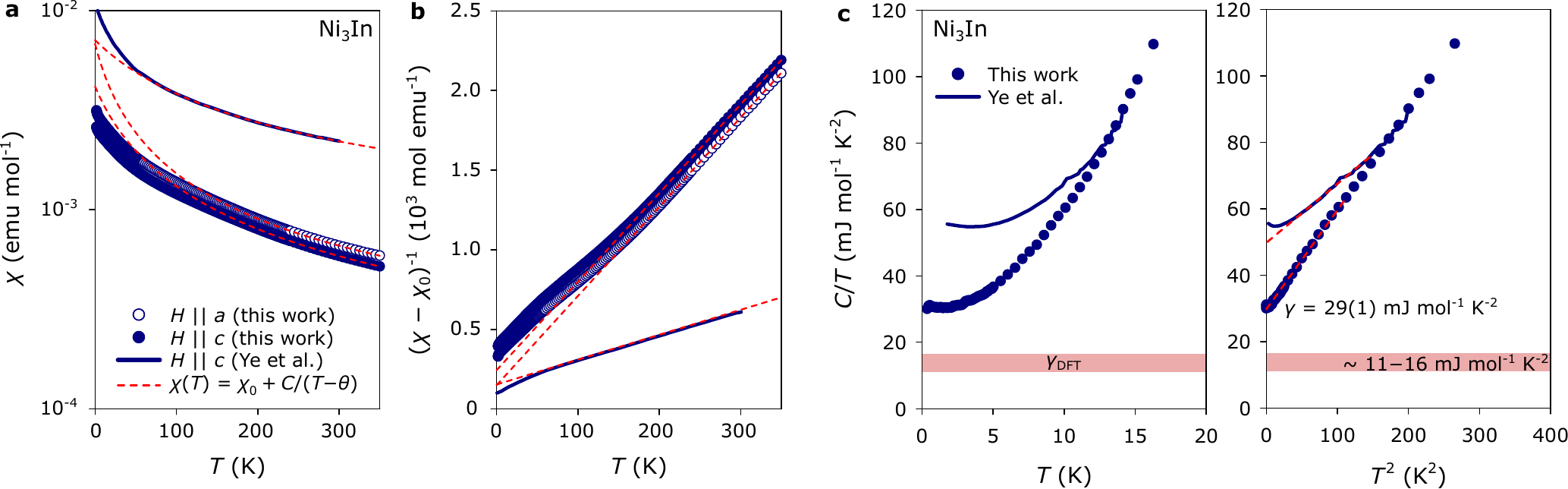}
\caption{\textbf{a} Temperature-dependent magnetic susceptibility of \ce{Ni3In} single crystals, measured with an applied field $\mu_0H = 0.1\,$T along the $a$- and $c$-axes. Red dashed lines represent least-squares fits to a modified Curie--Weiss law. For Pb-grown crystals, the data are well described in the temperature range 200\,--\,350\,K. In contrast, crystals grown by iodine-assisted vapor transport \cite{ye2024hopping} exhibit significantly larger $\chi(T)$ over the entire measured range and can be fitted down to 86\,K. From our fits, we obtain an effective moment of $\mu_\mathrm{eff} = 0.69\,\mu_\mathrm{B}$ per Ni for both $H \parallel a$ and $H \parallel c$, nearly a factor of two smaller than the value $\mu_\mathrm{eff} = 1.3\,\mu_\mathrm{B}$ reported for CVT-grown crystals \cite{ye2024hopping} as well as Curie--Weiss temperatures of $\theta_a \approx -27\,$K and $\theta_c \approx -44\,$K, which are also considerably smaller than those reported in Ref.\,\cite{ye2024hopping} ($\theta_a \approx -64\,$K and $\theta_c \approx -101\,$K). \textbf{b} Inverse magnetic susceptibility, highlighting the different high-temperature slope between crystals grown via different methods as well as minimal magnetic anisotropy in our samples. \textbf{c} Comparison of low-temperature specific heat $C/T$. The Sommerfeld coefficient $\gamma$ is about 29\,mJ\,mol$_\text{f.u.}^{-1}$\,K$^{-2}$, which is almost two times smaller than  $\gamma \approx 52$\,mJ\,mol$_\text{f.u.}^{-1}$\,K$^{-2}$ reported for the CVT-grown crystals \cite{ye2024hopping}, but still a factor of two to three larger than the Sommerfeld coefficient derived from the DFT density of states \cite{ye2024hopping,garmroudi2025topological}. The extracted value for the Debye temperature is about $\Theta_\text{D} \approx 297\,$K, slightly smaller than $\Theta_\text{D} \approx 327\,$K reported in Ref.\,\cite{ye2024hopping}.
}
\label{Fig4}
\end{figure*} 

\subsection{Magnetic and thermodynamic properties}
\label{magnetism}
Figure 4 summarizes the results on the magnetic susceptibility $\chi(T)$ and specific heat ($C/T$) of our \ce{Ni3In} single crystals, in comparison to CVT-grown samples reported previously in Ref.\,\cite{ye2024hopping}. It is evident that both the magnetic susceptibility (Fig.\,4a) and the specific heat (Fig.\,4c) of the Pb flux-grown crystals is substantially smaller than for the CVT-grown ones.

A modified Curie--Weiss (CW) fit,
\[
\chi(T)=\chi_0+\frac{C}{T-\theta},
\]
to our data yields an effective magnetic moment of $\mu_\mathrm{eff}\approx 0.69\,\mu_\mathrm{B}$ per Ni atom, independent of the field orientation, Curie--Weiss temperatures of $\theta_a \approx -27\,$K and $\theta_c \approx -44\,$K, and temperature-independent susceptibilities of $\chi_0=1.1\times10^{-4}$ and $6.2\times10^{-5}\,$emu\,mol$_\text{f.u.}^{-1}$ for $H\parallel a$ and $H\parallel c$, respectively. We note that the extracted value of $\mu_\mathrm{eff}$ is nearly a factor of two smaller than that reported in Ref.\,\cite{ye2024hopping}. Furthermore, the modified CW description is valid only above approximately $200\,\mathrm{K}$ in our samples, whereas the data obtained by Ye et al. \cite{ye2024hopping} could be fitted down to 86\,K.

We hypothesize that this discrepancy arises because $E_\text{F}$ is shifted further into the flat-band region, and thus deeper into the local-moment regime, in previously investigated CVT-grown samples. A glance at the binary Ni--In phase diagram reveals that \ce{Ni3In} is not a line compound, but instead exhibits a homogeneity range of \ce{Ni_{3+x}In_{1-x}} ($-0.02 < x < 0.02$), with elevated temperatures favoring slightly Ni-rich compositions \cite{durussel1997binary}. A small concentration of Ni antisite defects, which may shift the Fermi energy further into the flat-band states, could possibly account for the enhanced electrical resistivity, small or even negative magnetoresistance, and enhanced low-temperature specific heat (Fig.\,2a and Fig.\,4c). Indeed, the Sommerfeld coefficient of our samples, $\gamma \approx 29\,$mJ\,mol$_\text{f.u.}^{-1}$\,K$^{-2}$, is nearly a factor of two smaller than previously reported values, although it still exceeds the value estimated from the DFT density of states \cite{ye2024hopping,garmroudi2025topological}, $\gamma_\text{DFT} \approx 11$--$16\,$mJ\,mol$_\text{f.u.}^{-1}$\,K$^{-2}$, by roughly a factor of 2--3, highlighting the correlated nature of our \ce{Ni3In} crystals. We also note that the Kadowaki-Woods ratio $A/\gamma^2$ remains anomalously large in our crystals and is comparable to that reported previously, $A/\gamma^2 \approx 100\,\mu\Omega$\,cm\,K$^{2}$\,J$^{-2}$\,mol$_\text{f.u.}^{2}$.

Interestingly though, looking at the DFT-DOS in Fig.\,1b, it is obvious that a mere rigid-band-like shift of $E_\text{F}$ cannot quantitatively account for the discrepancy in $\gamma$ between our flux-grown crystals and the CVT-grown ones, since the DOS does not increase significantly below $E_\text{F}^\text{DFT}$. Similarly, we find that, while the resistivity increases with hole doping (Fig.\,S4), the maximum $\rho(T)$ obtained at realistic doping levels, is still significantly below the experimental values of Ref.\,\cite{ye2024hopping}. This suggests that (i) either a very small number of antisites substantially modify the DOS below $E_\text{F}$ not uncommon for transition metal-based systems \cite{qiu2010effect,sitaraman2021effect,parzer2022high,parzer2024semiconducting,parzer2025mapping} or (ii) an interesting scenario where correlation-driven effective mass renormalizations in the flat band may be highly doping-dependent. More studies, both on the experimental and theoretical side, are required to address this question.

\section{Conclusion}
\label{conclusion}
We have succeeded in growing millimeter-sized single crystals of the bilayer kagome metals \ce{Ni3(In,Sn)} using a Pb-flux method. Crystals synthesized in this manner exhibit markedly improved sample quality, as evidenced by significantly reduced residual resistivities, enhanced RRR, and a sizeable magnetoresistance compared to samples prepared via different techniques previously \cite{ye2024hopping,han2024molecular,garmroudi2024high}. The strange-metal-like behavior of the in-plane resistivity in \ce{Ni3In} persists over a similar temperature range, highlighting its intrinsic nature and indicating that it is not very sensitive to the growth technique. 

At the same time, the absolute values of $\rho(T)$, $\chi(T)$, and $C/T$ are consistently smaller by a factor of 2--3 in our crystals compared to previously reported samples grown by iodine-assisted recrystallization \cite{ye2024hopping}. We speculate that this discrepancy may possibly originate from Ni antisite defects, which could shift $E_\text{F}$ deeper into the flat-band regime.

\section{Acknowledgements}
Work at Los Alamos National Laboratory was performed under the auspices of the U.\,S. Department of Energy, Office of Basic Energy Sciences, Division of Materials Science and Engineering. F.\,G. acknowledges a
Director’s Postdoctoral Fellowship through the
Laboratory and Directed Research \& Development (LDRD) program. S.\,M.\,T. acknowledges support from the Los Alamos LDRD program.
The Flatiron Institute is a division of the Simons Foundation. 
Scanning electron microscopy and energy-dispersive x-ray measurements were performed at the Center for Integrated Nanotechnologies, an Office of Science User Facility operated for the U.S. Department of Energy (DOE) Office of Science.


%

\section{Experimental Methods}
\label{methods}
Single crystals of \ce{Ni3In_{1-x}Sn_x} were synthesized using a two-step procedure. First, polycrystalline \ce{Ni3In_{1-x}Sn_x} precursor ingots with a total mass of approximately 2\,g were prepared by arc melting stoichiometric amounts of elemental Ni (99.996\,\% purity), In (99.9995\,\%), and Sn (99.999\,\%). The resulting ingots were crushed into a coarse powder, mixed with Pb flux in a molar ratio of 1\,:\,25, and loaded into alumina crucibles, which were subsequently sealed under vacuum in quartz ampoules. 

The samples were heated to a maximum soak temperature $T_\text{soak}$ of either 800 or 1000\,$^\circ$C (see Table~1). After dwelling at $T_\text{soak}$ for 48\,h, the furnace was slowly cooled to 400\,$^\circ$C at a rate of 2\,$^\circ$C/h. The temperature was then raised again to 550\,$^\circ$C to facilitate removal of the Pb flux by centrifugation. To further remove residual flux, the crystals were carefully etched for a couple minutes in a mixture of glacial acetic acid and hydrogen peroxide. The obtained crystals were bar-shaped hexagonal prisms, with thicknesses of up to a few hundred micrometers and a length of up to a few millimeters, depending on the composition. Sn-rich compositions yielded noticeably thinner and smaller crystals, whereas \ce{Ni3In} grown from \ce{Ni3In_{0.5}Sn_{0.5}} yielded the largest crystals, as described and discussed in \autoref{growth}.

Single-crystal x-ray diffraction measurements were performed at room temperature using a Bruker D8 Venture diffractometer equipped with an Incotec I$\mu$S microfocus source employing Mo K$\alpha$ radiation ($\lambda = 0.71073$\,\AA). Diffraction data were collected with a PHOTON II CPAD detector and processed using the Bruker SAINT software package. Initial crystallographic models were obtained by intrinsic phasing methods implemented in SHELXT, followed by full-matrix least-squares refinements using SHELXL2018.

Energy-dispersive X-ray spectroscopy (EDX) and scanning electron microscopy (SEM) measurements were performed using a Scios 2 Dual Beam system by Thermo Fisher Scientific. Data were obtained for numerous spots on the same sample and for numerous samples of the same batch to confirm compositional homogeneity. The morphology and grain boundaries of the single-crystal surface were investigated in secondary electron emission mode, using an acceleration voltage of 20\,kV, a current of 1.6\,nA and a working distance of 6.8\,mm. The EDX results were analyzed using the Software Pathfinder.

Magnetic susceptibility measurements were carried out using the vibrating-sample magnetometer option of a Quantum Design Magnetic Property Measurement System (MPMS3). The magnetic susceptibility was measured in an applied field of 1\,kOe oriented both parallel and perpendicular to the crystallographic $c$ axis.

Heat-capacity measurements were performed using the two-$\tau$ thermal-relaxation method in a Quantum Design Physical Property Measurement System (PPMS). Measurements down to 0.37\,K were enabled by a $^3$He refrigerator insert. 

Electrical resistivity measurements between 1.8 and 300\,K were also carried out in the PPMS using a standard four-probe configuration with platinum wires attached to the sample surface using silver paint. Measurements down to 0.06\,K were performed in a Proteox dilution refrigerator.

\section{Computational Methods}

The DFT calculations of Ni$_3$In and Ni$_3$Sn used as a basis for the transport calculations reported in Fig~\ref{Fig1} were performed using \verb|Quantum ESPRESSO| v. 7.2 \sloppy \cite{quantumespresso_1, quantumespresso_2} with the GBRV pseudopotentials~\cite{garrity2014pseudopotentials} parameterized for the PBE exchange-correlation functional~\cite{perdew1996generalized} and using a 140\,Ry plane-wave coefficient energy cutoff. A 6$\times$6$\times$8 $k$ mesh and a 3$\times$3$\times$4 $q$ mesh were used for the initial coarse-grid density functional perturbation theory calculation of the electron-phonon matrix elements. 

Transport calculations were then performed with the Phoebe code~\cite{cepellotti2022phoebe}, an open-source package for Boltzmann transport equation solutions, using the relaxation time approximation (RTA) with an additional contribution from interband transitions included using the Wigner transport formalism.

Transport predictions were performed using a fine k-mesh sampling of the Brillouin zone, with a $110\times110\times145$ mesh for 20$-$200 K and a $83\times83\times110$ mesh for 200$-$300 K, to account for a finer sampling of the Fermi surface needed to converge transport calculations at low temperatures. Adaptive Gaussian smearing was used to broaden the delta functions found in the electron-phonon scattering rate expression.

\end{document}